\documentclass[12pt,a4]{article}
\usepackage{amsthm,amsmath,latexsym,amssymb,amsfonts,amscd}
\usepackage{graphics,lscape,fancyhdr,array,stmaryrd,euscript}
\usepackage{verbatim}
\usepackage{array}
\usepackage[all,cmtip]{xy}
\usepackage{color,tikz,tikz-cd}
\usetikzlibrary{snakes}

\usepackage[numbers,sort&compress]{natbib}
\usepackage[nottoc]{tocbibind}

\newcommand{\pl}{\partial}

\newcommand{\ga}{\alpha}
\newcommand{\gb}{\beta}
\newcommand{\gc}{\gamma}
\newcommand{\gd}{\delta}

\newcolumntype{w}[1]{%
>{\raggedright\hspace{0pt}}m{#1}}%
\newcolumntype{z}[1]{%
>{\raggedleft\hspace{0pt}}m{#1}}%

\newcommand{\besubeqs}{\begin{subequations}}
\newcommand{\esubeqs}{\end{subequations}}

\usepackage{amsthm,amsmath,latexsym,amssymb,amsfonts,amscd,hyperref}
\usepackage{color,tikz,tikz-cd}
\usepackage[all]{xy}

\theoremstyle{definition}

\theoremstyle{remark}

\numberwithin{equation}{section}

\begin{document}
\pagenumbering{gobble}
\hfill
\vskip 0.01\textheight
\begin{center}
{\Large\bfseries 
Deformation quantization of the simplest \\ Poisson Orbifold
\vspace{0.4cm}}

\vskip 0.03\textheight
\renewcommand{\thefootnote}{\fnsymbol{footnote}}
Alexey \textsc{Sharapov}${}^{a}$,  Evgeny \textsc{Skvortsov}\footnote{Research Associate of the Fund for Scientific Research -- FNRS, Belgium}${}^{b,c}$ \& Arseny \textsc{Sukhanov}${}^{d}$
\renewcommand{\thefootnote}{\arabic{footnote}}
\vskip 0.03\textheight

{\em ${}^{a}$Physics Faculty, Tomsk State University, \\Lenin ave. 36, Tomsk 634050, Russia}\\
\vspace*{5pt}
{\em ${}^{b}$ Service de Physique de l'Univers, Champs et Gravitation, \\ Universit\'e de Mons, 20 place du Parc, 7000 Mons, 
Belgium}\\
\vspace*{5pt}
{\em ${}^{c}$ Lebedev Institute of Physics, \\
Leninsky ave. 53, 119991 Moscow, Russia}\\
{\em ${}^{d}$ Moscow Institute of Physics and Technology, \\
Institutskiy per. 7, Dolgoprudnyi, 141700 Moscow region, Russia}

\end{center}

\vskip 0.02\textheight

\begin{abstract}
Whenever a given Poisson manifold is equipped with discrete symmetries the corresponding algebra of invariant functions or the algebra of functions twisted by the symmetry group can have new deformations, which are not captured by Kontsevich Formality. We consider the simplest example of this situation: $\mathbb{R}^2$ with the reflection symmetry $\mathbb{Z}_2$. The usual quantization leads to the Weyl algebra. While Weyl algebra is rigid, the algebra of even or twisted by $\mathbb{Z}_2$ functions has one more deformation, which was identified by Wigner and is related to Feigin's $gl_\lambda$ and to fuzzy sphere. With the help of homological perturbation theory we obtain explicit formula for the deformed product, the first order of which can be extracted from Shoikhet--Tsygan--Kontsevich formality. 
\end{abstract}
\newpage
\section{Introduction}
\pagenumbering{arabic}
\setcounter{page}{2}
The essence of the deformation quantization problem is to deform the usual point-wise product of smooth functions on a Poisson manifold $\mathcal{P}$; in so doing,  the first order deformation is supposed to be given by the Poisson bracket, $f\star g=fg+{i\hbar} \{f,g\}+\cdots$, and the problem is to complete the series in such a way that the new $\star$-product be associative. The solution to this problem was given by Kontsevich \cite{Kontsevich:1997vb} and was shown to result from the perturbative expansion of the path integral for the Poisson sigma-model \cite{Cattaneo:1999fm}. It was also proved in \cite{Kontsevich:1997vb} that non-equivalent quantizations of the same Poisson manifold are in one-to-one correspondence with formal (non-equivalent) deformations of the underlying Poisson bracket. In particular, if the Poisson bracket happens to be rigid (e.g. a symplectic manifold with trivial second de Rham cohomology), then the deformation quantization of $\mathcal{P}$ is essentially unique and the corresponding $\star$-product algebra, called the algebra of quantum observables, is rigid as well.

What if a given Poisson manifold $\mathcal{P}$ has some group $\Gamma$ of (discrete) symmetries? There are two more natural algebras in this situation. Instead of $C^\infty(\mathcal{P})$ one can consider the algebra $C^\infty(\mathcal{P})^\Gamma$ of $\Gamma$-invariant functions, that is, `smooth' functions on the orbifold $\mathcal{P}/\Gamma$. Alternatively, one can form the crossed product algebra $C^\infty(\mathcal{P}) \rtimes \Gamma$, extending $C^\infty(\mathcal{P})$ with the elements of $\Gamma$. A remarkable fact, see e.g. \cite{NeumaierPflaumPosthumaTang,Tiang,Halbout}, is that either algebra may admit new deformations! In particular, the quantized algebra $C^\infty(\mathcal{P})$ can have new non-trivial deformations when it is extended or quotiented by $\Gamma$. Therefore, there can be a few independent directions of quantization in the presence of symmetries, only one of which being captured by the Formality theorem.    

The general problem of deformation quantization of Poisson orbifolds is `easy' to solve: one just needs to quantize the Poisson sigma-model on a given Poisson orbifold, like it was done in \cite{Cattaneo:1999fm} for a Poisson manifold that is topologically $\mathbb{R}^{n}$. It is difficult, however, to give the general recipe leading to concrete expressions. It is also not clear if the problem can be solved in full generality or it is doomed to the case by case study. 

A rich class of examples of deformation quantization on Poisson orbifolds is provided by Symplectic Reflection Algebras \cite{EG}, \cite{2007arXiv0712.1568G}. These algebras were shown to implement the deformation quantization of the crossed product algebra $C^\infty(\mathbb{R}^{2n}) \rtimes \Gamma$, where $\mathbb{R}^{2n}$ is a linear symplectic space.  The proof given in \cite{EG} (see also \cite{Pinczon}) does not lead, however, to  explicit formulas, appealing to the PBW property of symplectic reflection algebras.  On the other hand, all the algebras $C^\infty(\mathbb{R}^{2n}) \rtimes \Gamma$ fall into the class captured by the general approach of \cite{Sharapov:2018hnl} that leads to explicit formulas for deformations. In this paper, we exemplify the approach by the simplest Poisson orbifold $\mathbb{R}^2/\mathbb{Z}_2$. Apart from Deformation Quantization, our interest is in the fact that this problem is related to the $3d$ bosonization duality in the large-$N$ limit, see \cite{Sharapov:2018kjz,Gerasimenko:2021sxj,Sharapov:2022eiy} and references therein. We first briefly review a number of different contexts where such an algebra showed up in the literature in Section \ref{sec:review} and present our results in Section \ref{sec:results}, which are supported by three technical Appendices.

\section{Deformed canonical commutation relations}\label{sec:review}
The simplest example of a Poisson orbifold appeared in the physics literature long ago \cite{Wigner1950}; since then it has reappeared in several other topics over the years, see e.g. \cite{Yang:1951pyq, Boulware1963, Gruber, Mukunda:1980fv,Vasiliev:1989re}. Wigner addressed the question to which extent the canonical commutation relations $[q,p]=i\hbar$ follow from the equations of motion in the case of a single particle in the harmonic potential: 
\begin{align}\label{eom}
    \tfrac{i}{\hbar}[H,q]&=p\,, & \tfrac{i}{\hbar}[H,p]&=-V'(q)\,, & H&= \tfrac12 p^2 +V(q)\,, && V(q)=\tfrac12 q^2\,.
\end{align}
He argued that since the Heisenberg equations of motion are more fundamental than the relation $[q,p]=i\hbar$ it makes sense to postulate the former and investigate to which extent they determine the latter. 
He found that $q,p$ should satisfy a weaker relation $([q,p]-i\hbar)^2=-u^2$ for some real constant $u$ (the notation is adapted to the discussion below). One can take the `square root' of this relation to get \cite{Yang:1951pyq, Boulware1963, Mukunda:1980fv}
\begin{align}\label{dqA}
    [q,p]&= i\hbar + i u R\,, && RqR=-q\,, && RpR=-p\,, &&R^2=1\,.
\end{align}
In other words, the quantum mechanical system is equipped with a parity operator\footnote{Similar operators appeared in the context of parastatistics and were called Klein operators, e.g. \cite{Luders,Druehl:1970fz,Schmutz,Ohnuki1982,Ohnuki:1984gz}.} $R$, which performs the reflection on $q,p$. Therefore, Wigner's result is that the canonical commutation relations can be deformed \eqref{dqA} without affecting \eqref{eom}. 

It is instructive to interpret this result from the deformation quantization vantage point. Let us start with the undeformed case, i.e., no $R$ and $u=0$. Classical observables form a commutative algebra $A_0={C}^\infty(\mathbb{R}^2)$ of functions $f(q,p)$ on the phase space. It is convenient to pack $q$, $p$ into a democratic two-component vector $y_\ga=\sqrt{2}(q\,,ip)$, $\ga,\gb,...=1,2$. As is well-known, the passage to the algebra of quantum observables, aka quantization, can be interpreted as a deformation of $A_0$ along the classical Poisson bracket. In other words, one is seeking for an associative product
\begin{align}
(f\star g)(y)&= (f\cdot g)(y) + \hbar \{f,g\}+\mathcal{O}(\hbar^2)=(f\cdot g)(y)+\sum_{k>0} \hbar^k B_k(f,g) 
\end{align}
that (i) starts with the usual point-wise multiplication on $A_0$; (ii) is given by a formal series in $\hbar$ with coefficients $B_k$ being bi-differential operators; (iii) the first-order deformation is determined by (or is equivalent to) the Poisson bracket. Up to equivalence, the general solution to this problem is provided by the Moyal--Weyl star-product:
\begin{align}
    (f\star g)(y)&= \exp \left[-\hbar\, \epsilon^{\ga\gb}\frac{\pl^2}{\pl y_1^\ga \pl y_2^\gb}\right]f(y_1)g(y_2)\Big|_{y_i=y}\,.
\end{align}
In particular, $[y_\ga, y_\gb]_\star=-2\hbar\epsilon_{\ga\gb}$.\footnote{A somewhat strange normalization here will pay back later.} This algebra is also known as the Weyl algebra $A_\hbar$. 
The Weyl algebra is rigid, meaning that it cannot be further deformed as an associative algebra. Therefore, the deformation found by Wigner relies crucially on adding a new element $R$, such that 
\begin{align}
    f(y,R)&:  && Ry_\ga R=-y_\ga\,, &&R^2=1\,.
\end{align}
We denote the algebra of functions $f(y,R)$ by $A_{0,0}$. Unlike $A_\hbar$, the algebra $A_{0,0}$ admits another non-trivial  deformation. 

In order to understand better what is going on it makes sense to review the general problem of deformation quantization (while in our case the Poisson bi-vector $\epsilon^{\ga\gb}$ is  constant and non-degenerate).
 An explicit construction of the deformed product on the initially commutative algebra $C^\infty(\mathcal{P})$ of functions on a Poisson manifold $\mathcal{P}$ is a  consequence of the Kontsevich formality theorem \cite{Kontsevich:1997vb}, of which the Moyal--Weyl star-product is an even simpler consequence. More generally, Poisson manifolds often come equipped with some (discrete) symmetries, e.g. reflections. Given a discrete group $\Gamma$ of symmetries of $\mathcal{P}$, one can be interested in two complementary algebras. First is the commutative algebra of $\Gamma$-invariant functions $C^\infty(\mathcal{P})^\Gamma$, which can also be viewed as the algebra of `smooth' functions on the orbifold $\mathcal{P}/\Gamma$. 
Second is the crossed product algebra $C^\infty(\mathcal{P}) \rtimes \Gamma$. The latter is generated by sums $\sum_i f_i \otimes \gamma_i$, where $f_i\in C^\infty(\mathcal{P})$ and $\gamma_i\in \Gamma$. The product is defined as $(f \otimes \gamma)\diamond (f' \otimes \gamma')=(f \gamma(f'), \gamma\gamma')$, where $\gamma(f)$ denotes the action of  $\gamma\in \Gamma$ on a function $f\in C^\infty(\mathcal{P})$. This algebra is already non-commutative even if $\Gamma$ is abelian.

Now, one can formulate a yet unsolved problem: how to quantize a general Poisson orbifold? Based on a number of simple cases \cite{AFLS,Tiang,Halbout}, one can argue that a given Poisson orbifold admits new deformations on top of the usual Kontsevich's one, the latter is always present. Wigner's example is the simplest instance of this additional deformation available. 

Coming back to the algebra $A_{0,0}$ of functions $f(y,R)$, which is slightly non-commutative due to $R$, we can treat  it as a crossed product $A_{0,0}=A_0\rtimes \mathbb{Z}_2$. There is also a closely related algebra $A_{0}^e=A_0/ \mathbb{Z}_2$ of even functions $f(y)=f(-y)$, i.e., functions on the orbifold $\mathbb{R}^2/\mathbb{Z}_2$. 

Now, let us summarize interpretations of various deformations of $A_0$, $A_0^e$ and $A_{0,0}$ since the deformations can also be understood in a number of different ways. 
\begin{description}
    \item[Algebra $\boldsymbol{A_0}$. ] Algebra $A_0=C^\infty(\mathbb{R}^2)$ admits a single deformation $A_\hbar$, which is along the Poisson bi-vector $\epsilon^{\ga\gb}$ and with a  parameter $\hbar$, the result being given by the Moyal--Weyl start product and is equivalent to the Weyl algebra.
    
    \item[Algebra $\boldsymbol{A_0^e}$. ] The orbifold algebra $C^\infty(\mathbb{R}^2)/\mathbb{Z}_2$ of even functions $f(y)$  enjoys two deformations. The first one is just the restriction $A_\hbar/\mathbb{Z}_2$ of the Weyl algebra $A_\hbar$ to even functions. In order to describe the second one, which is consistent with the first, it is useful to recall that bilinears $t_{\ga\gb}=-\tfrac1{2} y_\ga y_\gb$ form $sp_2$-algebra under the star-commutators 
    \begin{align}
        \tfrac{1}{\hbar}[t_{\ga\gb}, t_{\gc\gd}]_\star&= \epsilon_{\gb\gc} t_{\ga\gd}+\text{3 more}\,.
    \end{align}
    Since the even subalgebra of the Weyl algebra $A_\hbar$ can also be identified with functions $f(t_{\ga\gb})$ of $t_{\ga\gb}$, it has to be some quotient $U(sp_2)/I$ of the universal enveloping algebra $U(sp_2)$ of $sp_2$ by a certain two-sided ideal $I$. This ideal is generated \cite{Feigin} by $C_2-\text{const}$, where $C_2=-\tfrac12 t_{\ga\gb} t^{\ga\gb}$ is the Casimir operator of $sp_2$. In \cite{Feigin}, Feigin considered a one-parameter family of algebras $gl_\lambda =U(sp_2)/I_\lambda$, where $I_\lambda$ is generated by $C_2+\hbar^2(1-\lambda^2)$. For $\lambda=1/2$, $gl_\lambda$ is isomorphic to $A_\hbar$. It is this one-parameter family that corresponds to the second deformation of $A_\hbar^e$.  Feigin's $gl_\lambda$ is also closely related to the non-commutative sphere/hyperboloid, whose quantization was studied in \cite{Bieliavsky:2008mv}. Its name, $gl_\lambda$ originates from the fact that at $\lambda=\pm l$, $l=2,3,4,\ldots$, the algebra acquires a further two-sided ideal, the quotient being $gl(l)$.  Hence, $gl_\lambda$ interpolates between different matrix algebras. 
    
    \item[Algebra $\boldsymbol{A_{0,0}}$. ] The crossed product algebra $A_{0,0}$ of functions $f(y,R)$ admits the usual Moyal-Weyl deformation that turns it into $A_{\hbar,0}$. The product is defined by
    \begin{align}
        (f(y)+f'(y)R)\star(g(y)+g'(y)R)= f\star g+f'\star \tilde{g}'+(f\star g' +f'\star \tilde{g}) R\,,
    \end{align}
    where $\tilde{f}(y)\equiv f(-y)=Rf(y)R$ realizes the action of reflection  on $f$. Effectively, we find under the star-product that $[y_\ga, y_\gb]_\star=-2\hbar\epsilon_{\ga\gb}$. The algebra admits one more deformation $A_{\hbar,u}$ that can be described as the algebra of functions $f(q,R)$ in the generators $q_\ga$ obeying \cite{Yang:1951pyq, Boulware1963, Mukunda:1980fv}
    \begin{align}\label{dqC}
    [q_\ga,q_\gb]&= -\, 2\epsilon_{\ga\gb}(\hbar + u R)\,, && Rq_\ga R=-q_\ga\,, &&R^2=1\,.
    \end{align}
    $A_{\hbar,u}$ is a `parent' algebra for the other two algebras. Indeed, $A_\hbar=A_{\hbar,0}$. We can also restrict ourselves  to the even subalgebra $A^e_{\hbar,u}$, $f(q,R)=f(-q,R)$, which splits into two copies of $gl_\lambda$ with the help of idempotents $$\Pi_\pm=\tfrac12 (1\pm R)\,,\quad (\Pi_\pm)^2=\Pi_\pm\,,\quad \Pi_\pm\Pi_\mp=0\,,  \quad\,\Pi_++\Pi_-=1\,.$$ 
\end{description}

We would like to construct an explicit deformation quantization of $A_{0,0}$ that results in $A_{\hbar,u}$. It is remarkable that the algebra  $A_{\hbar,u}$ enjoys the PBW property \cite{EG}. Therefore, as a matter of principle, one can find out how to multiply two polynomials $f(q,R)$, $g(q,R)$ by repeatedly using Rels. \eqref{dqC}. We assume here, as always, that some ordering was chosen once and for all, otherwise one can multiply just by juxtaposition, $fg$, but it would be hard to tell if two differently looking expressions are the same or not unless an ordering is chosen. It is worth mentioning that the structure constants of $A_{\hbar,u}$ were obtained in the literature in a number of different forms \cite{Pope:1990kc,Bieliavsky:2008mv,Joung:2014qya,Korybut:2014jza,Basile:2016goq,korybut2020star}. This and more general algebras, known as symplectic reflection algebras, were extensively studied in the mathematical literature, see e.g. \cite{EG}, \cite{2007arXiv0712.1568G}. Our result is obtained via a different technique and reveals links to Shoikhet--Tsygan--Kontsevich formality \cite{Tsygan,Shoikhet:2000gw,FFS}.

\section{Quantization of the simplest Poisson Orbifold}\label{sec:results}
Our approach to deformation quantization of the simplest Poisson orbifold follows the idea of multiplicative resolutions \cite{Sharapov:2018hnl,Sharapov:2018ioy} and gives all bi-differential operators determining the deformation of the Moyal--Weyl star-product:
\begin{align}\label{prodexpanded}
(f\circ g)(y)&= (f\star g)(y) + u\, \phi_1(f,g) R+\mathcal{O}(u^2)=(f\star g)(y)+\sum_{k>0} u^k \phi_k(f,g) R^k \,,
\end{align}
$f$ and $g$ being polynomials in $y$'s. Here we explicitly indicate the dependence of  $R$. Let us note that any deformation of the crossed product algebra reduces to a certain deformation of the algebra of invariant functions. Therefore, we consider $A_{\hbar,0}$ as a starting point. It also makes sense to start with $A_{0,0}$ and consider a two-parameter deformation
\begin{align}
    f\circ g&= f\cdot g + \hbar \{f,g\} +u\{f,g\}_{\text{N.C.}}+\ldots \,,
\end{align}
where the second term is the usual Poisson bracket and the third one gives a certain non-commutative Poisson bracket which exists thanks to $R$.

The first order deformation $\phi_1(\bullet,\bullet)R$ has a special meaning. First of all, it is a Hochschild two-cocycle of $A_{\hbar,0}$, i.e., an element of $HH^2(A_{\hbar,0} ,A_{\hbar,0})$. Since $A_{\hbar,0}$ is the crossed product $A_\hbar \rtimes \mathbb{Z}_2$, it also has a clean interpretation as a Hochschild two-cocycle of the Weyl algebra with values in the dual module \cite{FFS}, that is, as an element of $HH^2(A_\hbar ,A^\star_\hbar)$. Indeed, if we identify the dual space  $A^\ast_\hbar$ with the Weyl algebra $A_\hbar$ itself by means of the non-degenerate inner product
\begin{align}
    g(f)= (f\star g)|_{y=0}\,, && f\in A_\hbar\,, \qquad g\in A_\hbar^*\,,
\end{align}
we can transfer the canonical bimodule structure of the Weyl algebra over itself -- $L_f(g)=f\star g$, $R_f(g)=-g\star f$ -- to its dual space to get $f\star g$ and $-g\star \tilde{f}$ for $f\in A_\hbar$ and $g\in A_\hbar^*$. A Hochschild two-cocycle $\phi_1$ of $HH^2(A_\hbar ,A^\star_\hbar)$ is then a nontrivial solution to
\begin{align} \label{Ideform}
    a\star\phi_1(b,c)-\phi_1(a\star b,c)+\phi_1(a,b\star c)-\phi_1(a,b)\star\tilde{c}&=0\,, &&a,b,c\in A_\hbar\,.
\end{align}
Extending $A_\hbar$ with $R$ to get $A_{\hbar,0}$ allows one to represent both the adjoint and the dual actions of $A_\hbar$ as the adjoint action of $A_{\hbar,0}$ on itself. Therefore, any non-trivial cocycle of $A_{\hbar,0}$ can be represented as $\phi_1R$.

While the existence of a non-trivial $\phi_1$ is easy to see \cite{Feigin1983,Feigin1989,FFS}, its explicit form emerges \cite{FFS} from Shoikhet--Tsygan--Kontsevich formality theorem \cite{Shoikhet:2000gw}. In fact, it gives a non-trivial $2n$-cocycle for each $A^{n}$, where $A^{n}$ is the Weyl algebra in $n$ pairs of generators. The construction of all $\phi_k$ below is such that it does reproduce $\phi_1$ in its original form \cite{FFS}.  

\subsection{Deformation via a resolution}\label{S3.1}
The multiplicative resolution is based on the extension of algebra $A_{\hbar,0}=C^\infty(\mathbb{R}^2)\rtimes \mathbb{Z}_2$ with further coordinates $z_\ga$ and associated anti-commuting differentials $dz^\ga$.\footnote{For this particular case this construction is very similar to the one of \cite{Vasiliev:1990cm}. } The additional relations are $Rz_\ga R=-z_\ga$, $Rdz^\ga R=-dz^\ga$. This algebra is equipped with a variant of the star-product that is defined via its symbol (it suffices to define it on $R$- and $dz$-independent functions)
\begin{align}\label{superstar}
   m(f,g)\equiv (f\ast g)(y,z)&=\exp[\hbar p_{12} +2 p_1\cdot q_2 ]f(y+y_1,z+z_1)g(y+y_2,z+z_2)\Big|_{y_i=z_i=0}\,,
\end{align}
where we used a shorthand notation for symbols of poly-differential operators $y\equiv p_0$, $\pl_{y_1}\equiv p_1$, $\pl_{y_2}\equiv p_2$, $q_1\equiv\pl_{z_1}$, $q_2\equiv \pl_{z_2}$; indices are contracted $q\cdot p \equiv q^\ga p_\ga$ in such a way that $\exp[p_0 \cdot p_1]f(y_1)= f(y+y_1)$; $p_{ij}\equiv p_i \cdot p_j$; we often omit the arguments and the projection $|_{y_i,z_i=0}$. 
In terms of the generators we find
\begin{align*}
    y_\ga\ast f&= (y_\ga -\hbar \pl^y_\ga -2\pl^z_\ga)f\,,  & z_\ga\ast f&= z_\ga f\,,\\
    f\ast  y_\ga &= (y_\ga +\hbar \pl^y_\ga)f\,,  & f\ast z_\ga &= (z_\ga +2\pl^y_\ga)f \,.
\end{align*}
The deformed product \eqref{prodexpanded} is obtained via homological perturbation theory (HPT), see Appendices \ref{app:HPT} and \ref{app:unfoldingHPT}. In few words, the initial algebra $A_{\hbar,0}$ can be identified with the cohomology of the exterior differential $d_z=dz^\ga \pl^z_{\ga}$ in $z$-space. We then introduce a specific deformation of this differential, where a key role is played by a closed central two-form $\lambda=\exp[z\cdot y] R \,dz_\alpha dz^\alpha$.   The homological perturbation theory generates a certain deformed product. Technically, $\phi_k$ are given by all trees with two leaves, we call $a$, $b$, representing the arguments of $\phi_k(a,b)$ and $k$ leaves corresponding to a special element (a gauge potential of a certain magnetic field in $z$-space):
\begin{align}
    A&= dz^\nu z_\nu  \int_0^1 t \,dt\,  \exp[t z\cdot y] R = h[ dz^2 \exp[z\cdot y] R]=h[\lambda]\,.
\end{align}
Here, $h$ is the standard contracting homotopy for the de Rham complex in $z$-space:
\begin{align}
    h[dz^2 f(z)]&= dz^\ga z_\ga \int_0^1 t \,dt\, f(zt)\,, &
    h[dz^\ga g_\ga]&= z^\ga \int_0^1 \,dt\, g_\ga(zt)\,,
\end{align}
and we complete the definition by setting $h[q(z)]=0$ for any zero-form $q(z)$. The trees have only one type of a vertex --- a trivalent one corresponding to $m$ (two arguments in, one out).

\paragraph{First order.} There are four trees one can draw at the lowest order and only one of them survives to give a non-vanishing contribution thanks to our choice of the star-product and of the homotopy:\footnote{ See Appendix \ref{app:unfoldingHPT} for more detail. In short: all graphs that have $h[...]$, i.e. $h$ as the last operation vanish upon $z=0$ projection; all graphs that have $h (...\ast a)$ vanish because $z^\ga z_\ga=0$.}
$$
   \phi_1(a,b)=a(y) \ast h[ b(y) \ast A ]= \begin{tikzcd}[column sep=small,row sep=small]
   & {}& \\
    & m\arrow[u]  & \\
    a\arrow[ur]  & & m\arrow[ul, "h" ']   & \\
    & b\arrow[ur]& &A\arrow[ul]
\end{tikzcd}
$$
This graph gives
\begin{align}
    \phi_1(a,b)&=4p_{12}  \int t \exp[p_{01}(1-2 tt')+p_{02}(1-2t)+\hbar\, p_{12}(1-2t+2tt')]\,,
\end{align}
where $t$ and $t'$ should be integrated over $[0,1]$. One can reduce the integral over the square to the one over the $2d$ simplex $\Delta_2=\{0<u_1<u_2<1\}$:
\begin{align}
    \phi_1(a,b)&=4p_{12} \int_{\Delta_2} \exp[p_{01}(1- 2u_1)+p_{02}(1-2u_2)+\hbar\, p_{12}(1-2u_2+2u_1)]\,.
\end{align}
Here, we explicitly see propagators $G(u,v)=1+2(u-v)$ that emerge from  Shoikhet-Tsygan-Kontsevich Formality, see \cite{FFS}, so that
\begin{align}
    \phi_1(a,b)&=4p_{12} \int_{\Delta_2} \exp[p_{01} G(0,u_1)+p_{02}G(0,u_2)+\hbar\, p_{12}G(u_1,u_2)]\,.
\end{align}
The integral can be taken to give \cite{FFS} 
($p_{01}=x$, $p_{02}=y$, $p_{12}=z$ and with $\hbar=1$)
\begin{align}
   \phi_{1}&= \frac{z e^{-x-y+z}}{(x+y) (x-z)}+\frac{z e^{x+y+z}}{(x+y) (y+z)}-\frac{z e^{x-y-z}}{(x-z) (y+z)}\,.
\end{align}
In terms of symbols the cocycle condition for $\phi_1$ reads
\begin{align}\notag
    \phi_1(p_0+\hbar p_1,p_2,p_3) e^{p_{01}}-\phi_1(p_0,p_1+p_2,p_3) e^{\hbar p_{12}}+\phi_1(p_0,p_1,p_2+p_3) e^{\hbar p_{23}}-\phi_1(\hbar p_3+p_0,p_1,p_2) e^{-p_{03}}=0\,.
\end{align}
Associated to this Hochschild two-cocycle there is also a cyclic cocycle, see \cite{Sharapov:2020quq} for an explicit formula. 

\paragraph{Second order.} At the second order there is again only a single tree that contributes:
$$
   a(y) \ast h[ h[ b(y) \ast A ] \ast A]= \begin{tikzcd}[column sep=small,row sep=small]
   &{}&\\
    & \arrow[u] m  & \\
    a\arrow[ur]&&  \arrow[ul,"h"']m &  \\
    & \arrow[ur,"h"]m &&\arrow[ul]A \\
    \arrow[ur]b & & \arrow[ul]A &
\end{tikzcd}
$$
which, after a straightforward algebra, gives
\begin{align*}
    \phi_2(a,b)=16p_{12}^2\int \left(1-2 t_1\right) t_1 t_3 t_4 \exp \Big[& p_{01} (-2 t_1 t_2 t_4+4 t_1 t_2 t_3 t_4-2 t_3 t_4+1) +\\
    +&p_{02} (4 t_3 t_1-2 t_1-2 t_3+1)+ \\
    +&\hbar\, p_{12}(4 t_3 t_1+2 t_2 t_4 t_1-4 t_3 t_4 t_1-2 t_1-2 t_3+2 t_3 t_4+1)\Big]\,.
\end{align*}
Here, all times $t_{1,2,3,4}$ should be integrated over $[0,1]$. 
The associativity of the product \eqref{prodexpanded} breaks down to (mind the reflection automorphism on one of the arguments):
\begin{align*}
        -a&\star \phi_2(b,c)+\phi_2(a\star b,c) -\phi_2(a,b\star c) +\phi_2(a,b)\star c- \phi_1(\phi_1(a,b),\tilde{c})+\phi_1(a,\phi_1(b,c))=0\,.
\end{align*}
It is the associativity of $a\star b +\phi_1(a,b)R+\phi_2(a,b)+...$ to the second order and $R$ results in $\tilde{c}$. We can also simplify the integral a bit by extracting a $2d$-simplex out of the integration domain, see below. 

\paragraph{General formula.} It can be shown that the parameters of the resolution are fine-tuned in such a way that with the standard homotopy $h[...]$ only a single tree survives at each order to give  
\begin{align}
    \phi_k(a,b)&= a\ast h[ h[\cdots  h[b\ast A]\ast A\cdots]\ast A ]\Big|_{z=0}\,.
\end{align}
It is easy to get a formula for any $n$, see Appendix \ref{app:unfoldingHPT} for a proof. After some change of integration variables as to reveal the simplex associated with the nested homotopy operation we find 
\begin{align*}
    \phi_n(f,g)&= \phi_n(p_0,p_1,p_2)  f(y_1)\,g( y_2)\Big|_{y_i=0}\,,\\
    \phi_n&=4^n \int_{B_n} \prod _{j=1}^n v_j \left(1-2 v_j\right){}^{n-j} \times (p_{12})^n\exp\Big[p_{01} a_{01}+p_{02} a_{02}+\hbar\, p_{12}a_{12}\Big]\,,\\
    a_{01}&=1-2 \sum _{j=1}^n u_j v_j \prod _{k=j+1}^n \left(1-2 v_k\right)\,,\\
    a_{02}&=\prod _{j=1}^n \left(1-2 v_j\right)\,,\\
    a_{12}&=a_{02}+2 \sum _{j=1}^n u_j v_j \prod _{k=1}^{j-1} \left(1-2 v_k\right)\,,
\end{align*}
where the integration is over $B_n$ that is a product of the $n$-simplex $\Delta_n=\{0<u_1<u_2<...<u_n<1\}$ and $n$-cube $v_i \in [0,1]$. It is important to remember about $R$, which is present as the rightmost overall factor $\phi_n(a,b) R^n$. The formula here-above is one of the main results of this paper. 

We note that the exponent is linear in the simplex variables, which was the main reason to introduce them. Just as an example for $n=1$ we have
\begin{align*}
    a_{01}&=1-2 u_1 v_1\,, &
    a_{02}&=1-2 v_1\,, &
    a_{12}&=a_{02}+2 u_1 v_1\,.
\end{align*}
In this case one can join $1$-simplex, $0<u_1<1$, and the interval $v_1\in [0,1]$ (which are the same) into a $2$-simplex, as we did for $\phi_1$ above. 

Another comment is that the prefactor $(p_{12})^n$ tells us that as the order $n$ increases the deformation does not affect polynomials of order less than $n$, which can be easily seen from the deformed commutation relations \eqref{dqC}. $v_i$ appear in a more and more non-linear way, while the exponent is always linear in $u_i$. Therefore, the $u$-integral can be explicitly evaluated, if needed. There is a simple formula that does that, see below. It  might also be useful to replace $v_j$ with $w_i=(1-2v_i)\in[-1,1]$, then
\begin{align*}
    \phi_n&= \int_{C_n} \prod _{j=1}^n (1-w_j) \left( w_j\right){}^{n-j} \times (p_{12})^n\exp\Big[p_{01} a_{01}+p_{02} a_{02}+\hbar \, p_{12}a_{12}\Big]\,,\\
    a_{01}&=1- \sum _{j=1}^n u_j (1-w_j) \prod _{k=j+1}^n w_k\,,\\
    a_{02}&=\prod _{j=1}^n w_j\,,\\
    a_{12}&=a_{02}+ \sum _{j=1}^n u_j (1-w_j) \prod _{k=1}^{j-1} w_k\,,
\end{align*}
where $C_n =\Delta_n \times [-1,1]^n$.

\paragraph{Weyl deformation switched off, Dunkl derivative. } We can set $\hbar=0$ and get an interesting deformation of $A_{0,0}$ along the reflection operator $R$ direction, which leads to $A_{0,u}$. In this limit, the deformation can also be realized \cite{TianDunkl} via the Dunkl derivative \cite{Dunkl}. However, a polarization has to be chosen for $y_\ga$ to do that, which breaks manifest $sp_2$-symmetry that $\phi_k$ enjoy. On the other hand, the Dunkl derivative is quite simple, while infinitely many $\phi_k$ are needed to maintain $sp_2$-invariance. It would be interesting to see if, at least for $\hbar=0$, $\phi_k$ can be resummed. Following \cite{TianDunkl}, one introduces complex coordinates $w$ and $\bar{w}$ instead of $y_\ga$ to write
\begin{align}
    f\circ g&= f g + \frac{\hbar}{2} \frac{f(w,\bar w)-f(-w,\bar w)}{2w}\frac{g(-w,\bar w)-g(-w,-\bar w)}{2\bar  w}R\,.
\end{align}
At least for $\phi_1$ the commutative limit gives a simple expression after evaluating the integral (it is non-singular despite its appearance, recall that $p_{01}=x$, $p_{02}=y$, $p_{12}=z$)
\begin{align}
   \phi&= z \left(\frac{e^{-x-y}}{x (x+y)}+\frac{e^{x+y}}{ y (x+y)}-\frac{e^{x-y}}{ x y}\right)\,,
\end{align}
which obeys ($a\cdot b$ is the point-wise product, which is commutative)
\begin{align*}
    -a&\cdot \phi_1(b,c)+\phi_1(a\cdot b,c) -\phi_1(a,b\cdot c) +\phi_1(a,b)\cdot \tilde{c}=0\,,
\end{align*}
or in terms of symbols
\begin{align}
    \phi_1(p_0,p_2,p_3) e^{p_{01}}-\phi_1(p_0,p_1+p_2,p_3) +\phi_1(p_0,p_1,p_2+p_3) -\phi_1(p_0,p_1,p_2) e^{-p_{03}}=0\,.
\end{align}

\paragraph{Localization formula for an integral over a simplex.} Suppose that we need to integrate $\exp[f]$ over the $n$-dimensional simplex $\Delta_n$:
\begin{align*}
    \Omega=\int_{\Delta_n}\exp[f]\, du_1\wedge ...\wedge du_n\,.
\end{align*}
Suppose now that $f=\sum_i a_i u_i$ is some linear form. Then, the integral is given by the following localization formula
\begin{align}
   \int_{\Delta_1^n} \Omega&= \sum_{v\in V} \exp\Big[f\big|_{v}\Big] \times \left( \prod_{e\in E(v)} \nabla_{\vec{e}}f\right)^{-1}\,,
\end{align}
i.e. it is a sum over the vertices $v\in V$ with the exponent given by its value at a vertex $v$ times the inverse of the product of directional derivatives with the respect to all the edges $e$ meeting at this vertex. Since every body can be cut into/approximated by the simplest simplices, the range of application of this formula can be broader.

\paragraph{Result for general $\boldsymbol{A}$.} Let us assume that we have some functions of $(y_i,z_i)$ $i=3,4,...$ instead of our specific $A$. We also assume that these functions are of the form $z_\nu dz^\nu\exp[y\cdot p_i +z\cdot q_i] f(y_i,z_i)$, $i=3,4,...$, i.e. we have some one-forms and, hence, exactly the same trees will contribute as for $\phi_k$. Then, 
the general formula for what can be denoted $\phi_n(a,b,f_3,...,f_n)$ reads
\begin{align*}
    \phi_n&= 4^n \prod _{i=2}^{n+1} \sum _{j=2}^i p_1\cdot p_j \exp\Big[\sum _{i=0}^{n+2} \sum _{j=i+1}^{n+2} p_i\cdot p_j+2 \sum _{j=1}^n u_j p_1\cdot q_{j+2}+2 \sum _{i=2}^{n+2} \sum _{j=i+1}^{n+2} p_i\cdot q_j\Big]\,,
\end{align*}
where $u_i$ belong to the simplex. Now we can apply it to $f_j= \exp[t_j z_j \cdot y_j]R$ to get $\phi_k$ quoted above. The change of variables that is required reads (there are $2n$ times $t_i$ at order $n$):
\begin{align*}
    t_{2k-1}&= v_{n+1-k}\,, &
    t_{2n}&= u_1\,, &
    t_{2n-2k}&=u_{k+1}/u_k \,.
\end{align*}
This result can be useful if one wishes to represent $\phi_k$ as a correlation function in some auxiliary quantum mechanics (similarly to how the Poisson sigma model that implements \cite{Cattaneo:1999fm} Kontsevich's formula reduces to a quantum mechanics on a circle for a non-degenerate Poisson structure). The results already obtained can be extended further to get a certain $A_\infty/L_\infty$-algebra, see Appendix \ref{app:hisgra}.

\section*{Acknowledgments}
\label{sec:Aknowledgements}
We would like to thank Xiang Tang for a very useful correspondence. The work of E.S. was partially supported by the European Research Council (ERC) under the European Union’s Horizon 2020 research and innovation programme (grant agreement No 101002551) and by the Fonds de la Recherche Scientifique --- FNRS under Grant No. F.4544.21. A. Sh. gratefully acknowledges the financial support of the Foundation for the Advancement of Theoretical Physics and Mathematics “BASIS”. The work of A.S was supported by the Russian Science Foundation grant 18-72-10123 in association with the Lebedev Physical Institute. 

\appendix

\section{HPT and deformation}
\label{app:HPT}

Let $(V,d_V)$ and $(W, d_W)$ be a pair of complexes.   A 
{\it strong deformation retract} (SDR) associated to them is given by  chain
maps $p:V\rightarrow W$ and $i:W\rightarrow V$ such that $pi=1_W$
and $ip$ is homotopic to $1_V$. The last property implies the
existence of a map $h: V\rightarrow V$ such that
$$
dh+hd=ip-1_V\,.
$$
Without loss in generality, one may also assume the following {\it
annihilation properties}:
$$
hi=0\,,\qquad ph=0\,,\qquad h^2=0\,.
$$
All these data are  summarized by the following diagram:
\begin{equation}\label{SDR}
\xymatrix{ *{\hspace{5ex}(V,d_V)\;}\ar@(ul,dl)[]_{h}
\ar@<0.5ex>[r]^-p & (W, d_W) \ar@<0.5ex>[l]^-i}\,.
\end{equation}
A particular case of SDR is $W=H(V)$ and $d_W=0$.

\vspace{3mm}\noindent {\bf Perturbation Lemma} \cite{Brown65}.
{\it Given SDR data (\ref{SDR}) and a small perturbation $\delta$ of
$d_V$ such that $(d_V+\delta)^2=0$ and $1-\delta h$ is invertible,
one can define a new SDR
$$
\xymatrix{ *{\hspace{9ex}(V,d_V+\delta )\;}\ar@(ul,dl)[]_{h'}
\ar@<0.5ex>[r]^-{p'} & (W, d'_W) \ar@<0.5ex>[l]^-{i'}}\,,
$$
where the maps are given by
$$
\begin{array}{ll}
     p'=p+p(1-\delta h)^{-1}\delta h\,,&\quad i'=i+h(1-\delta h)^{-1}\delta i\,, \\[3mm]
    h'=h+h(1-\delta h)^{-1}\delta h\,, & \quad d'_W = d_W+p(1-\delta h)^{-1}\delta i\,.
\end{array}
$$
}

In many practical cases one can think of the operator $(1-\delta h)^{-1}$ as being defined
by the geometric series
$$
(1-\delta h)^{-1}=\sum_{n=0}^\infty (\delta h)^n\,,
$$
whose convergence is ensured by a suitable filtration in $V$.

Among basic applications of homological perturbation theory are the transference and deformation problems for various algebraic structures. Here we are concerned with the structure of an associative algebra on $V$.  Therefore, consider an associative dg-algebra $(A, d)$ over a ground field $k$ with differential $d: A\rightarrow A$ and product $m: A\otimes A\rightarrow A$. Regarding $A[1]$ as just a graded vector space,\footnote{By definition, $A[1]_n=A_{n+1}$. On shifting degree by one unit, the product $m$ acquires degree $1$, while $d$ retains its degree $1$.} we can also define the tensor coalgebra $T^cA[1]$ for the
Alexander--Whitney coproduct
$$
\Delta: T^cA[1]\rightarrow T^cA[1]\otimes T^cA[1]\,.
$$
$$
\Delta (a_1\otimes \cdots\otimes a_n)=1\otimes (a_1\otimes\cdots
\otimes a_n)+\sum_{i=1}^{n-1}(a_1\otimes\cdots\otimes a_i)\otimes
(a_{i+1}\otimes\cdots\otimes a_n)$$
$$
+(a_1\otimes\cdots \otimes a_n)\otimes 1\,.
$$
The coproduct is coassociative in the sense that  $\Delta (\Delta\otimes 1)=\Delta(1\otimes \Delta)$. 
The differential  and the product in $A$ extend then canonically\footnote{Recall that a  linear map $F: T^cA[1]\rightarrow T^cA[1]$ is called a {coderivation}, if it obeys the co-Leibniz rule $\Delta F=(F\otimes 1+1\otimes F)\Delta$.
It is well known (see e.g. \cite{jones2002lectures}) that the space of coderivations is isomorphic to  $\mathrm{Hom}_k(TA[1],A[1])$, so that  any homomorphism $f: TA[1]\rightarrow A[1]$ defines and is defined by a coderivation $\hat{f}: T^cA[1]\rightarrow T^cA[1]$. Notice that $T^0A[1]=k$ and $f: T^0A[1]\rightarrow A[1]$ is determined by an element $\lambda\in A[1]$.} to coderivations on $T^cA[1]$: 
$$
\hat d(a_1\otimes \cdots\otimes a_n)=\sum_{k=1}^n \pm a_1\otimes \cdots \otimes da_k\otimes \cdots \otimes a_n\,,
$$
$$
\hat m(a_1\otimes \cdots\otimes a_n)=\sum_{k=1}^{n-1} \pm a_1\otimes \cdots \otimes m(a_k, a_{k+1})\otimes \cdots \otimes a_n\,.
$$
Here $\pm$ is the usual Koszule sign. Notice that $\hat d$ defines the canonical differentials in the tensor powers of the complex $A$. It is straightforward to check that both the coderivations square to zero and commute to each other, i.e., $\hat d^2=0$, $\hat m^2=0$, and $[\hat d,\hat m]=0$. Hence, either of them makes $T^cA[1]$ into a codifferential coalgebra. 
Let $H(A)$ denote the cohomology algebra of the dg-algebra $(A,d)$.

\vspace{3mm}\noindent {\bf Tensor Trick} \cite{gugenheim1989, gugenheim1991perturbation}. 
{\it With any SDR 
$$
\xymatrix{ *{\hspace{5ex}(A,d)\;}\ar@(ul,dl)[]_{h}
\ar@<0.5ex>[r]^-p & (H(A), 0) \ar@<0.5ex>[l]^-i}
$$
one can associate new SDR data
\begin{equation}\label{SDR2}
\xymatrix{{(\,}\ar@(ul,dl)[]_-{\hat{h}} &\ar@<0.5ex>[r]^-{\hat{p}}{ \hspace{-6ex} T^cA[1], \hat d
)\;} & (T^cH(A)[1],
0) \ar@<0.5ex>[l]^-{\hat{i}}}\,,
\end{equation}
where 
$$
\hat{p}=\sum_{n=1}^\infty p^{\otimes n}\,,\qquad \hat{i}=
\sum_{n=1}^\infty i^{\otimes n}\,,\qquad 
\hat{h}=\sum_{n\geq 1}^\infty \sum_{k=0}^{n-1} 1^{\otimes k}\otimes
h\otimes (ip)^{\otimes n-k-1}\,.
$$
}
The operator $\hat h$ is called sometimes the Eilenberg--Maclane homotopy.

Let us apply the Perturbation Lemma above to the following situation: 
$$V=T^c A[1]\, , \qquad d_V=\hat d \,,\qquad \delta=\hat m\,,\qquad  W=T^c H(A)[1]\,, \qquad  d_W=0\,.$$
This gives immediately  a new SDR 
\begin{equation}\label{SDR2}
\xymatrix{ {(\,}\ar@(ul,dl)[]_-{\hat{h}}&{\hspace{-6ex}T^cA, \hat d+\hat m
)\;} \ar@<0.5ex>[r]^-{\hat{p}'} & (T^cH(A),
\hat{\mu}) \ar@<0.5ex>[l]^-{\hat{i}'}}
\end{equation}
with 
\begin{equation}
\hat{\mu} = \hat p(1-\hat m\hat h)^{-1}\hat m \hat i=\sum_{k=0}^\infty \hat p(\hat m\hat h)^k\hat m\hat i\,.
\end{equation}
Suppose further that $i(H(A))$ is a subalgebra in $A$. This assumptions holds e.g. for the algebra of Sec. \ref{S3.1}. Then $\hat h\hat m\hat i=0$ and the last formula simplifies to $\hat \mu =\hat p \hat m \hat i $, where $\mu$ is  nothing but the associative product in $H(A)$ induced by $m$ in $A$. In such a way we `solved' the transference problem for the associative product $m$ in $A$. The result is the associative product $\mu$ in the `smaller' space $H(A)$.  

Let us now turn to the deformation problem. For this end, we take any central cocycle  of the dg-algebra $\lambda$ of $A$, that is, 
\begin{equation}\label{lambda}\lambda\in Z(A)\,,\qquad d\lambda=0\,.\end{equation}
This gives rise to a coderivation 
$\hat \lambda : T^cA[1]\rightarrow T^cA[1]$ that acts by the rule 
\begin{equation}
\begin{array}{c}
    \hat\lambda (a_1\otimes \cdots\otimes a_n)=\lambda\otimes a_1\otimes\cdots\otimes a_n\\[2mm]
    \displaystyle+\sum_{k=1}^{n-1}\pm a_1\otimes \cdots \otimes a_{k}\otimes {\lambda} \otimes a_{k+1}\otimes \cdots \otimes a_n\pm a_1\otimes\cdots\otimes a_n\otimes \lambda\,. 
    \end{array}
\end{equation}
With this $\hat \lambda $ we can perturb the perturbation $\hat m$ above to get a new perturbation $\delta=\hat m+\hat \lambda$. It is clear that $(\hat d+\hat m+\hat \lambda)^2=0$ as a consequence of (\ref{lambda}). Applying the the Perturbation Lemma once again, we obtain the SDR 
\begin{equation}\label{SDR2}
\xymatrix{ {(\,}\ar@(ul,dl)[]_-{\hat{h}}&{\hspace{-6ex}T^cA, \hat d+\hat m+\hat \lambda
)\;} \ar@<0.5ex>[r]^-{\hat{p}'} & (T^cH(A),
\hat{\mu}') \ar@<0.5ex>[l]^-{\hat{i}'}}\,,
\end{equation}
where
\begin{equation}
\hat{\mu}' = \hat p(1-(\hat m+\hat \lambda)\hat h)^{-1}(\hat m+\hat \lambda ) \hat i\,.
\end{equation}
If $\hat h \hat m\hat i=0$, then the last formula simplifies to
\begin{equation}
\hat{\mu}' = \hat \mu+\hat p(1-(\hat m+\hat \lambda)\hat h)^{-1}\hat \lambda  \hat i\,,
\end{equation}
$\mu =p m i $ being the product in $H(A)$. 
This gives a deformed associative product in cohomology:
$$
\mu'(a, b)=\mu(a, b)+\hat p(1-(\hat m+\hat \lambda)\hat h)^{-1}\hat \lambda\hat i(a\otimes b)\qquad \forall a, b\in 
H(A)\,.
$$
It is convenient to depict the non-deformed product $\mu(a,b)$ in $H(A)$ by the vertex graph
$$
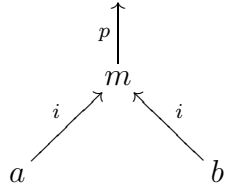
\xymatrix{
&&\\
&m\ar[u]^p&\\
a\ar[ur]^i&&b\ar[ul]_i
}
$$
Then the first-order deformation $\mu_1(a,b)$ is described by the sum of four graphs:
$$
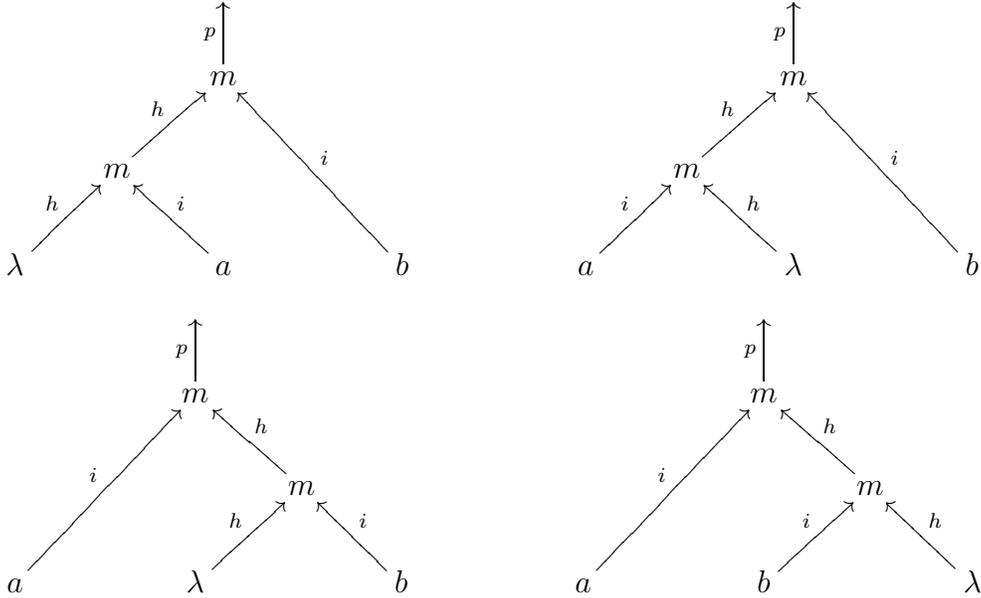
\begin{array}{cc}
\xymatrix{
&&&&\\
&& m\ar[u]^p&&\\
&m\ar[ur]^h&&&\\
\lambda\ar[ur]^h&&a\ar[ul]_i&&b\ar[uull]_i
}
&\qquad \qquad
\xymatrix{
&&&&\\
&& m\ar[u]^p&&\\
&m\ar[ur]^h&&&\\
a\ar[ur]^i&&\lambda\ar[ul]_h&&b\ar[uull]_i
}
\\
\xymatrix{
&&&&\\
&& m\ar[u]^p&&\\
&&&m\ar[ul]_h&\\
a\ar[uurr]^i&&\lambda\ar[ur]^h&&b\ar[ul]_i
}
&\qquad\qquad
\xymatrix{
&&&&\\
&& m\ar[u]^p&&\\
&&&m\ar[ul]_h&\\
a\ar[uurr]^i&&b\ar[ur]^i&&\lambda\ar[ul]_h
}
\end{array}
$$
In the main text and from now on we will depict trees in a simplified form, omitting the obvious projections $i$, $p$. The second-order deformation is given by the relation 
\begin{equation}
    \mu_2(a,b)=\hat p(\hat m \hat h\hat \lambda\hat h\hat m\hat h\hat m\hat h+\hat m \hat h\hat m \hat h\hat \lambda \hat h\hat m\hat h+\hat m\hat h\hat m \hat h\hat m\hat h\hat \lambda\hat h) \hat\lambda\hat i(a\otimes b)\,.
\end{equation}
As is seen, the only non-zero products are those in which the number of $\lambda$'s is one less than the number of $m$'s and the number of $h$'s is twice the number of $\lambda$'s. These rules hold true in all higher orders. 

In our specific situation the data for the HPT are: (a) differential $d=d_z=dz^\ga \pl^z_{\ga}$ is the exterior differential in $z$-space; (b) multiplication $m$ is defined by \eqref{superstar} on functions $f(y,z,dz,R)$; (c) deformation $\lambda=\exp[z\cdot y] R \,dz_\alpha dz^\alpha$ is a central, $d_z$-closed two-form.

\section{Unfolding HPT}
\label{app:unfoldingHPT}
After the general procedure is explained in Appendix \ref{app:HPT}, we would like to identify the trees that contribute to $\phi_k$ and evaluate them explicitly. The star-product \eqref{superstar} leads to the following particular cases, which we need in practice,
\begin{align*}
    g(y,z)\ast f(y) &= \exp[yp_i] g(y-p_i,z) f(y_i)\big|_{y_i=0}\,,\\
    f(y)\ast g(y,z)&= \exp[yp_i] g(y+p_i,z+2p_i) f(y_i)\big|_{y_i=0}\,,
\end{align*}
where $f(y)$ and $g(y,z)$ are two arbitrary functions of their arguments. The arguments of $\phi_k(a,b)$ are functions of $y$ only.

We can start building a tree from the bottom of some branch. We cannot have $A$'s as the two leaves at the bottom since $h[A\ast A]=0$, as can be verified. Therefore, we have either $b\ast A$ or $A\ast b$ ($a\ast b$ will not lead to a non-vanishing result after we try to add $A$'s). We have to apply $h$ immediately, otherwise we cannot build a tree that is zero-form. Now, we note that $h[A\ast b]=0$ due to $z^\ga z_\ga\equiv0$. Indeed, the star-product on $y-z$ is chosen to discriminate between left and right arguments and $A\ast b$ is still proportional to $dz^\ga z_\ga$, while $b\ast A$ has a $z$-independent term as well. Therefore, there is only one way to grow a non-trivial branch: one starts with $h[b\ast A]$ and applies $h[X \ast A]$ to the result $X$ of the previous step; at the end one has to multiply by $a$ from the left. This way we find a single contribution
\begin{align}
    \phi_k(a,b)&= a\ast h[ h[... h[b\ast A]\ast A...]\ast A ]\Big|_{z=0}\,.
\end{align}
Since the result must be $z$-independent, one can set $z$ to any value. Our choice of a projection is $z=0$. Now, we derive a recurrence relation for the coefficients in the exponent of $\phi_k$ and for the prefactor. We first write

\begin{align}
    h&\left(h\left(...h\left(b*A\right)...*A\right)*A\right)=2^{n}\int_0^1d t_1 t_1\int_0^1d t_2 t_2\ \dots\int_0^1d t_n t_n\times\nonumber\\
    \times&\int_0^1d k_n\ \dots\int_0^{k_3}d k_2\int_0^{k_2}d k_1 \ b\left(y_2\right) \left(z^{\beta}p_{2\beta}\right)\left(z^{\beta}p_{3\beta}\right)\dots\left(z^{\beta}p_{n+1\beta}\right)\times\nonumber\\
    \times&\exp\left(y_3^{\nu}p_{2\nu}\left(1-2t_1\right)-t_1k_1y_3^{\nu}z_{\nu}+t_1k_1z^{\nu}p_{2\nu}\right)\times\nonumber\\
    \times&\exp\left(y_4^{\nu}p_{3\nu}\left(1-2t_2\right)-t_2k_2y_4^{\nu}z_{\nu}+t_2k_2z^{\nu}p_{3\nu}\right)\times\nonumber\\
    &\dots\nonumber\\
    \times&\exp\left(y_{n+2}^{\nu}p_{n+1\nu}\left(1-2t_n\right)-t_n k_n y_{n+2}^{\nu}z_{\nu}+t_n k_n z^{\nu}p_{n+1\nu}\right)=...\,,\nonumber
\end{align}
where a number of auxiliary $y_i$, $i=3,...,n+2$ was introduced and we set $y^{\nu}_{n+2} \equiv p^{\nu}_0$. We also made the simplex part of the domain of integration explicit. Now we can isolate the last step
\begin{align}
   ... =&2^n\int_0^1d t_1 t_1\int_0^1d t_2 t_2\ \dots\int_0^1d t_n t_n \int_0^1d k_n\ \dots\int_0^{k_3}d k_2\int_0^{k_2}d k_1\times\nonumber\\
    \times&b\left(y_2\right)\omega_n = 2^n\int_0^1d t_1 t_1\int_0^1d t_2 t_2\ \dots\int_0^1d t_n t_n\times\nonumber\\
    \times&\int_0^1d k_n\ \dots\int_0^{k_3}d k_2\int_0^{k_2}d k_1 b\left(y_2\right)\left(z^{\beta}p_{n+1\beta}\right)\times\nonumber\\
    \times&\exp\left(y_{n+2}^{\nu}p_{n+1\nu}\left(1-2t_n\right)-t_n k_n y_{n+2}^{\nu}z_{\nu}+t_n k_n z^{\nu}p_{n+1\nu}\right)\omega_{n-1}\label{h(...*A)^n2}\,.
\end{align}
Assuming that $\omega_{i-1}=\left(z^{\beta}p_{2\beta}\right)^{i-1}c_{i-1}\exp\left(y_{i+1}^{\nu}p_{2\nu}d_{i-1}- y_{i+1}^{\nu}z_{\nu}e_{i-1}+z^{\nu}p_{2\nu}f_{i-1}\right)$ we find
\begin{align}
    \omega_i&=\left(z^{\beta}p_{i+1\beta}\right)\exp\left(y_{i+2}^{\nu}p_{y+1\nu}d'_{i}- y_{i+2}^{\nu}z_{\nu}e'_{i}+z^{\nu}p_{i+1\nu}f'_{i}\right)\omega_{i-1}=\nonumber\\
    =&c_{i-1}\left(z^{\beta}p_{2\beta}\right)^{i-1}\left(z^{\beta}p_{i+1\beta}\right)\exp\left(y_{i+2}^{\nu}p_{y+1\nu}d'_{i}- y_{i+2}^{\nu}z_{\nu}e'_{i}+z^{\nu}p_{i+1\nu}f'_{i}\right)\times\nonumber\\
    \times&\exp\left(y_{i+1}^{\nu}p_{2\nu}d_{i-1}- y_{i+1}^{\nu}z_{\nu}e_{i-1}+z^{\nu}p_{2\nu}f_{i-1}\right)=d_{i-1}c_{i-1}\left(z^{\beta}p_{2\beta}\right)^i\times\nonumber\\
    \times&\exp\left(y_{i+2}^{\nu}p_{2\nu}\left(d_{i-1}d'_{i}\right)- y_{i+2}^{\nu}z_{\nu}\left(e_{i-1}d'_i+ e'_i\right)+z^{\nu}p_{2\nu}\left(f'_{i}d_{i-1}+f_{i-1}\right)\right)=\nonumber\\
    =&c_i\left(z^{\beta}p_{2\beta}\right)^i\exp\left(y_{i+2}^{\nu}p_{2\nu}d_{i}- y_{i+2}^{\nu}z_{\nu}e_{i}+z^{\nu}p_{2\nu}f_{i}\right)\,.
\end{align}
The initial data is given by
\begin{equation}
   \omega_1 = \left(z^{\beta}p_{2\beta}\right)\exp\left(y_3^{\nu}p_{2\nu}\left(1-2t_1\right)-t_1k_1y_3^{\nu}z_{\nu}+t_1k_1z^{\nu}p_{2\nu}\right) \,,
\end{equation}
and, therefore, we get for any $i$
\begin{equation}
     \omega_i=c_i\left(z^{\beta}p_{2\beta}\right)^i\exp\left(y_{i+2}^{\nu}p_{2\nu}d_{i}- y_{i+2}^{\nu}z_{\nu}e_{i}+z^{\nu}p_{2\nu}f_{i}\right)\label{omega_i}\,,
\end{equation}
where
\begin{align}
    &c_1=1\,,\quad&&d_1=\left(1-2t_1\right)\,,\quad&&e_1=t_1k_1\,,\quad &&f_1=t_1k_1\nonumber\,,\\
    &c'_i=1\,,&&d'_i=\left(1-2t_i\right)\,,&&e'_i=t_ik_i\,, &&f'_i=t_ik_i\,,\nonumber\\
    &c_i=d_{i-1}c_{i-1}\,,&&d_i=d_{i-1}d'_i\,,&&e_i = e_{i-1}d'_i+ e'_i\,, &&f_i = f'_{i}d_{i-1}+f_{i-1}\,.\label{coof}
\end{align}
The latter recurrence relations are easy to solve 
\begin{align}
    &d_i=\prod_{j=1}^i\left(1-2t_j\right)\label{coof1}\,,\\
    &e_i=\sum_{j=1}^it_j k_j\prod_{l=j+1}^{i}\left(1-2t_l\right),\ \mbox{where}\ \prod_{l=i+1}^{i}\left(1-2t_l\right)=1\label{coof2}\,,\\
    &f_i=\sum_{j=1}^it_j k_j\prod_{l=1}^{j-1}\left(1-2t_l\right),\ \mbox{where}\ \prod_{l=1}^0\left(1-2t_l\right)=1\label{coof3}\,,\\
    &c_i=\prod_{j=1}^{i-1}\left(1-2t_j\right)^{i-j},\ \mbox{where}\ \prod_{j=1}^{0}\left(1-2t_j\right)^{i-j}=1\,,\label{coof4}
\end{align}
and the solution is given in the main text.

\section{Strong Homotopy Algebra}
\label{app:hisgra}
Given the general result for any $A$, we can substitute $f=h[dz^2C(y)\ast \exp[ z\cdot y]R]$ to get an $A_\infty/L_\infty$-algebra of the formal higher spin gravity \cite{Vasiliev:1990cm}. It does not correspond to physically acceptable interactions \cite{Boulanger:2015ova}, but is still well-defined as an $A_\infty/L_\infty$-algebra. For more detail we refer to \cite{Sharapov:2019vyd}, where it is explained that any one-parameter family of associative algebras can be used to construct a certain $A_\infty$-algebra and the associated $L_\infty$. Following the same steps as for $\phi_k$ we find for the non-trivial structure maps
\begin{align}
    m_n(a,b,c_1,...,c_n)&=2^{2n}\int_0^1d t_1 t_1\int_0^1d t_2 t_2\ \dots\int_0^1d t_n t_n \times\nonumber\\
    \times&\int_0^1d k_n\ \dots\int_0^{k_3}d k_2\int_0^{k_2}d k_1\ \prod_{j=1}^{n-1}\left(1-2t_j\right)^{n-j}\left(p_{12}\right)^n\times\nonumber\\
    \times&\exp\left[\prod_{j=1}^n\left(1-2t_j\right)p_{02}+\left(1-2\sum_{j=1}^n t_j k_j\prod_{l=j+1}^{n}\left(1-2t_l\right)\right)p_{01}+\right.\nonumber\\
    +&\left.\left(2\sum_{j=1}^n t_j k_j\prod_{l=1}^{j-1}\left(1-2t_l\right)+\prod_{j=1}^n\left(1-2t_j\right)\right)p_{12}+\right.\nonumber\\
    +&\left.\sum_{i=1}^n\left(2t_i k_i -4 t_i\sum_{j=1}^{i-1}t_jk_j\prod_{l=j+1}^i\left(1-2t_l\right)\right)p_{1,i+2}+\right.\nonumber\\
    +&\left.\sum_{i=1}^n\left(2t_i\prod_{j=1}^{i-1}\left(1-2t_j\right)\right)p_{2,i+2}\right]a\left(y_1\right)b\left(y_2\right)c_1\left(y_3\right)\dots c_n\left(y_{n+2}\right)\,,
\end{align}
where we took out all $R$ factors and assumed that the empty products equal $1$ $$\prod_{l=i+1}^{i}\left(1-2t_l\right)=\prod_{l=1}^0\left(1-2t_l\right)=1\,.$$

\footnotesize
\providecommand{\href}[2]{#2}\begingroup\raggedright\endgroup

\end{document}